\documentclass[]{emulateapj}
\usepackage{subfigure,rotating,graphics,soul,array}
\usepackage{epstopdf}

\def\deg{\hbox{$^\circ$}}              
\def\lum{$\rm{erg}~\rm{s}^{-1}$}       
\def\lx{$\rm{L}_{X}$}                   
\def\zsun{$\rm{Z}_{\odot}$}           

\def\td{$\tau_{d}$}
\def\tr{$\tau_{r}$}

\def\xmm{{\it XMM-Newton}}

\slugcomment{The Astronomical Journal}

\shorttitle{X-ray flare on 47 Cas}
\shortauthors{Pandey \& Karmakar}

\begin{document}

\title{An X-ray flare from 47 Cas}

\author{Jeewan C. Pandey and  Subhajeet Karmakar}
\affil{Aryabhatta Research Institute of Observational Sciences (ARIES), 
Nainital-263 002, India}
\email{jeewan@aries.res.in}

\begin{abstract}
Using \xmm ~observation, we investigate properties of a flare from the very active and poorly known stellar system 47 Cas. The luminosity at the peak of the flare was found to be $3.54\times 10^{30}$ \lum, which is $\sim 2$ times more than that at quiescent state. The quiescent state corona of 47 Cas was represented by two temperature plasma: 3.7 and 11.0 MK.    
 The time-resolved X-ray spectroscopy of the flare showed the variable nature of the temperature, the emission measure,  and the abundance. The maximum temperature during the flare was found to be 72.8 MK.  We inferred the length of a flaring loop to be   $3.3 \times 10^{10}$ cm using a hydrodynamic loop model. Using the RGS spectra, the density during the flare was found to be $4.0\times 10^{10}$ cm$^{-3}$.   The loop scaling laws were also applied when deriving physical parameters of the flaring plasma.

\end{abstract}


\section{Introduction}
X-ray flares are common phenomena in late-type stars and it is  believed that they are produced by the same mechanism as solar flares, namely magnetic reconnection. However, stellar flares show very high temperatures and an extremely large total energy. Observationally, stellar flares are defined  as a  sudden enhancement of an intensity followed by a gradual decay.
The study of stellar flares is a valuable tool for understanding stellar coronae as these are dynamical events that  consist of different information than the quiescent state observations \citep{favata05,getman08a,getman08b,pandey08,pandey12}.

\begin{figure*}
\begin{center}
\subfigure[PN]{\includegraphics[width=75mm]{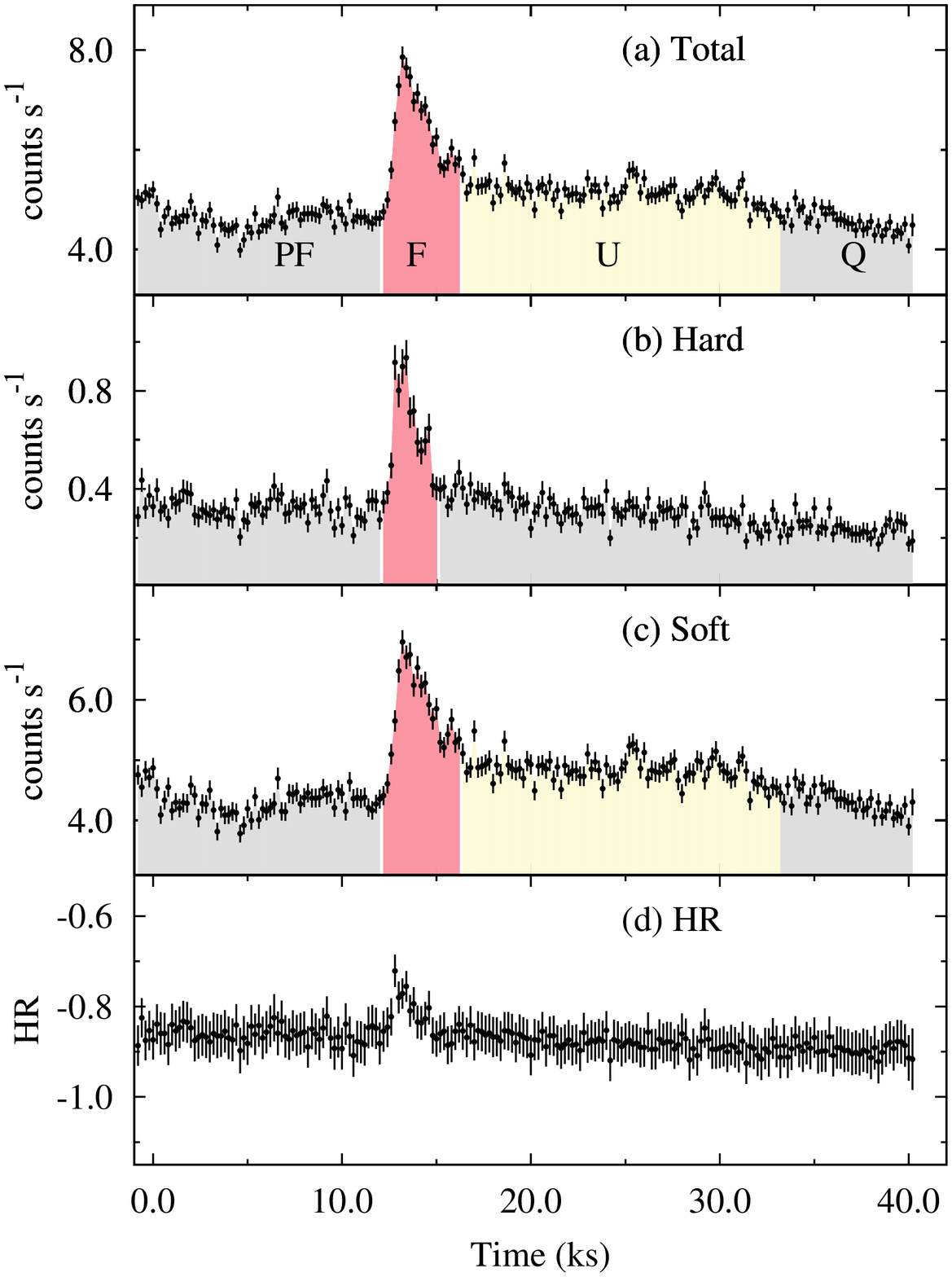}}
\hspace{0.2cm}
\subfigure[MOS]{\includegraphics[width=75mm]{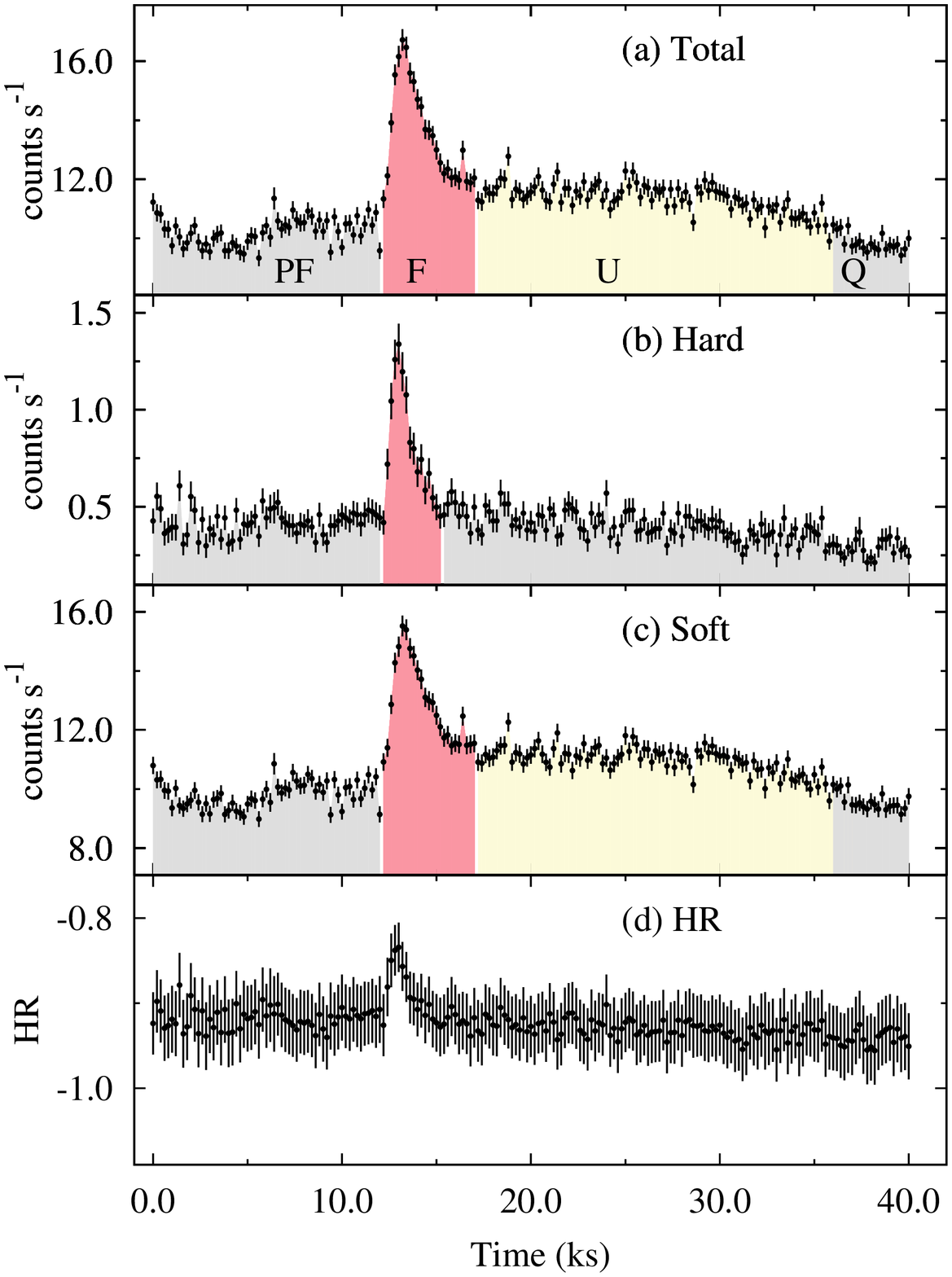}}
\caption{X-ray light curve of the 47 Cas system at three different energy bands along with the hardness ratio curve. The hardness ratio is defined as $(H-S)/(H+S)$. The pre-flare state, flaring state, heightened post-flare emission and  quiescent state are  marked by PF,  F, U, and Q, respectively.}
\label{fig:lc}
\end{center}
\end{figure*}

For the present study, we have chosen a poorly known early F-type main sequence stellar system, 47 Cas.  It is supposed that the 47 Cas system consists of an unseen companion, 47 Cas B, which has been detected only in the radio wavelength by \cite{guedel98} and for which no optical characterization  is available  thus far. The HIPPARCOS catalog lists  the young and rapidly rotating F0V star, 47 Cas, as a close visual binary with  a period of about 1616 days. \cite{guedel95, guedel98} have studied 47 Cas using X-ray (ROSAT)  and radio (6cm) observations. Their study suggests that X-ray and radio emissions from 47 Cas are due to the late-type companion.  \cite{garner98}  have also suggested that the X-Ray and radio activities  are similar to a chromospherically active solar-type companion. Furthermore,  signatures of coronal activity are not normally associated with early F-type main-sequence stars. Later, the high resolution X-ray spectra of 47 Cas from \xmm ~were analyzed by \cite{ness03,telleschi05} and \cite{nordon08} with an aim of abundance analysis.  In this paper, we have analyzed the flaring feature in the 47 Cas  system observed from \xmm. The paper is organized as follows: observations and data reduction are described in Section \ref{sec:obs},   analyses of \xmm ~observations are explained in Section  \ref{sec:analysis},  and loop parameters are derived in Section \ref{sec:loop}. Finally, we  discuss and conclude our results in Section \ref{sec:disc}. 

\section{Observations and data reduction}
\label{sec:obs}
47 Cas was observed by the \xmm ~satellite using the European Photon Imaging Camera (EPIC) and Reflection Grating Spectrometer (RGS) instruments on 2001 September 11 at 02:21:19 UT for 40 ks.  The EPIC consists of three CCDs behind three X-ray telescopes \citep{jansen01}; the two identical metal oxide semiconductor (MOS) CCDs, MOS1 and MOS2  \citep{turner01}, and one {\it p-n} junction  CCD, PN \citep{struder01}. The EPIC instrument provides imaging and spectroscopy in the energy range of 0.15\textendash 15 keV with an angular resolution of 4.5-6.6 arcsec and a spectral resolution ($E/\Delta E$) of 20-50.  The RGS consists of two identical grating spectrometers, RGS1 and RGS2, behind different mirrors \citep{denherder01}. The dispersed photons are recorded by a strip of eight CCD MOS chips. One of these chips has failed in each spectrometer, leading to gaps in the spectra which fortunately affect different spectral regions. The RGS covers the wavelength range of approximately 6\textendash 38 \AA ~({\textit E} = 2.5\textendash 0.3 keV) with a resolution of 0.05 \AA ~and a peak effective area of about 140 cm$^2$ at 15 \AA.

We have used standard \xmm ~Science Analysis System (SAS) software, version 13.5.0  with updated calibration files for reducing the EPIC and RGS data of 47 Cas. The pipeline processing of raw EPIC Observation Data Files was done using the {\sc epchain} and {\sc emchain} tasks, which allow calibrations both in energy and astrometry of the events registered in each CCD chip. However, the metatask {\sc rgsproc} was used to generate the RGS event files. For the EPIC data, we have restricted our analysis to the energy band 0.3\textendash 10.0 keV due to background contribution at high energies where  stellar sources have undetectable fluxes.  We have also checked the data for pileup and high background proton flares and they are free from these effects.  The SAS task {\sc evselect} was used to extract the event list files.  X-ray light curves and spectra from the EPIC observations  of the 47 Cas system were generated from on-source counts obtained from circular regions with a radius of 40 arcsec  around the source. The background was chosen from source-free regions on the detectors surrounding the source.  In order to  correct the light curve for good time intervals, dead time, exposure, point-spread function, quantum efficiency, and background contribution,  the SAS {\sc epiclccorr}  was used . However, the EPIC spectra were generated from  {\sc especget}, which also computes the photon redistribution as well as the ancillary matrix. Finally, the spectra were rebinned to have a minimum of 20 counts per spectral bin.

\section{Analysis and results}
\label{sec:analysis}

\subsection{X-ray light curves}
\label{sec:lightcurve}
Figure \ref{fig:lc} shows the background subtracted  X-ray light curve of 47 Cas in the total (0.3\textendash 10.0 kev), soft (0.3\textendash2.0 keV) and hard (2.0\textendash 10.0 kev) energy bands with a temporal binning of 200s. The light curves in all energy bands show variability,  which resembles flaring activity.  The flaring feature is  marked with an `F' in Figure \ref{fig:lc}.  The flare began after 11.2 ks from the start of PN observations and lasted for 4.8 ks.  In the total energy  band, the count rate at the peak of the flare was found to be 1.8 times more than that in the quiescent state. However, the flare peak to quiescent state count ratios were found to be 1.7 and 4.4 in the soft and hard energy bands, respectively. After the end of the flare an active level `U' was identified where the average flux was 1.2 times more than that in the quiescent  state.  The level `U' was identified only in the soft and total energy bands, and no such feature was seen in the hard energy band.  In the total energy band,  e-folding rise (\tr) and decay (\td) times of the flare observed in 47 Cas  were derived to be $831\pm100$ s and $2494\pm 82$ s, respectively. However, \tr ~and  \td ~were derived as $743\pm89$ s and $2446\pm77$ s, and $590\pm108$ s and $1203\pm62$ s in the soft and hard energy bands, respectively. 

The bottom panel of Figure \ref{fig:lc} shows the temporal  variation of the hardness ratio (HR). The HR is defined as $ (H-S)/(H+S)$; where $H$ and $S$ are the count rates in the hard and soft bands, respectively. The variation in the HR during the flares is indicative of changes in the coronal temperature.  The HR varied in a  fashion similar to its light curves, indicating an increase in the temperature at the flare peak and a subsequent cooling. 
\subsection{X-ray Spectral Analysis}

\subsubsection{Quiescent state spectra}
The quiescent state X-ray spectra were extracted from the `Q' part of the light curve. The quiescent state coronal parameters of 47 Cas  were derived by fitting the X-ray  spectra with a single (1T) and  double (2T) temperature collisional plasma models known as {\sc apec} \citep{smith01} as implemented in the X-ray spectral fitting package {\sc xspec} \citep{arnaud96} version 12.8.1.  The global abundances ($Z$) and interstellar hydrogen column density ($N_H$) were left as free parameters.  The $N_H$ is modeled with cross sections obtained by \cite{morrison83}; however, the solar photospheric  abundances ($Z_\odot$) were adopted from \cite{anders89}.  Both 1T and 2T plasma models with solar photospheric abundances  were rejected due to the high value of $\chi^2$. The 2T model with with  sub-solar abundances was found to be acceptable with a reduced  $\chi^2$ of 1.2.  By adding one more component to the  temperature in fitting, we did not find any further improvement in the reduced  $\chi^2$; therefore, we assume that the quiescent coronae of 47 Cas were well represented by two temperatures plasma. The cool and hot temperatures and  the corresponding emission measures were derived as  $0.32\pm 0.02$ keV and $0.95\pm0.02$ keV, and  $9.0_{-1.2}^{+1.3}\times10^{52}$ cm$^{-3}$ and $17.7_{-1.3}^{+1.4}\times10^{52}$ cm$^{-3}$, respectively. The value of $N_H$ was derived to be $1.6\pm0.7 \times 10^{20}$ cm$^{-2}$, which is  lower than that of the total galactic {\sc Hi} column density \citep{dickey90} towards the direction of 47 Cas. The abundance during the quiescent state of 47 Cas was derived to be $0.135\pm 0.009 Z_\odot$   The quiescent state X-ray flux of  47 Cas  was estimated by using the {\sc cflux} model in {\sc xspec} and are corrected for $N_H$. The unabsorbed X-ray luminosity of 47 Cas during its quiescence was derived to be $1.90\pm0.02 \times 10^{30}$ \lum.  

\subsubsection{Flare Spectra}
\begin{figure}
\begin{center}
\includegraphics[angle=-90,width=80mm]{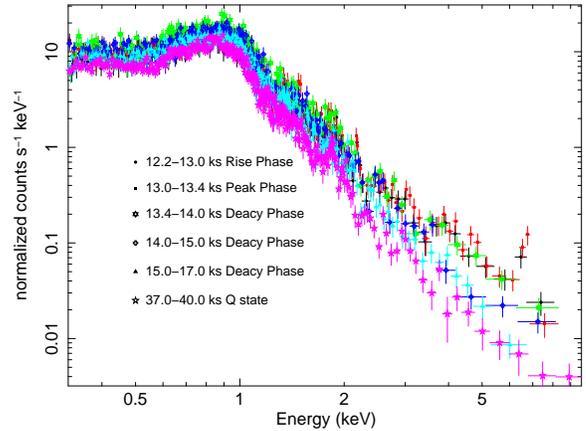}
\caption{X-ray spectral evolution of 47 Cas during the flare with respect to quiescent state.}
\label{fig:spec}
\end{center}
\end{figure}

\begin{figure}[t]
\begin{center}
\includegraphics[width=80mm]{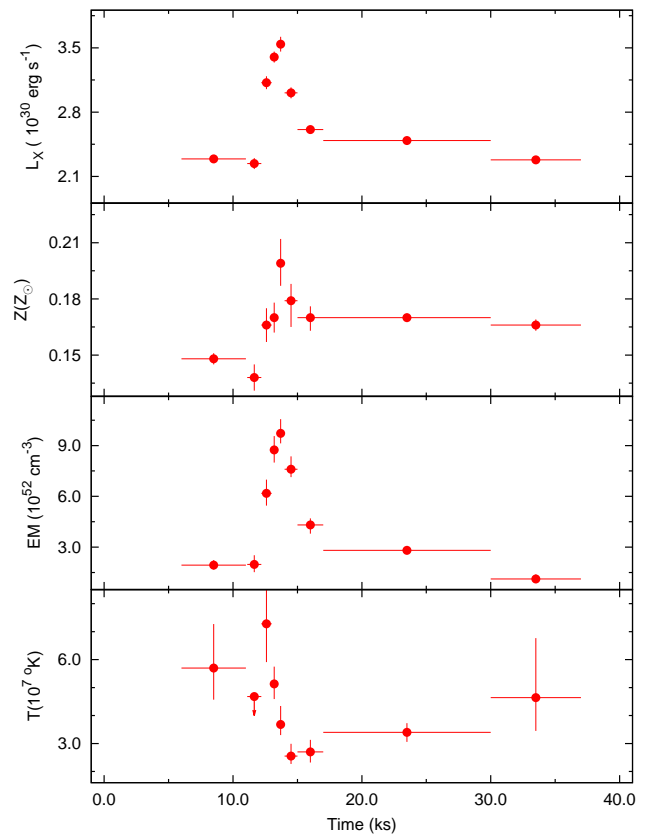}
\caption{Evolution of X-ray spectral parameters of 47 Cas during the flare.}
\label{fig:para}
\end{center}
\end{figure}

The X-ray light curve of 47 Cas  was divided into  11 time bins, and spectra of each time interval were extracted and analyzed to trace the spectral changes during the flare.  These divisions  were made to ensure that spectra in each time bin have sufficient counts to provide reliable values of spectral parameters. Figure \ref{fig:spec} shows the spectra of different time intervals during the flaring and quiescent states. The spectral evolution can be seen  clearly. In order to study the flare emission only, we have performed three temperature spectral fits of the data, with the
quiescent emission taken into account by including its best-fitting 2T model as a frozen background contribution, which  allows us to derive one "effective'' temperature and one EM of the flaring plasma. Initially, $N_H$ was a free parameter in the spectral fitting and appeared to be  constant during the flare to the quiescent state value. Therefore, in the next stage of spectral fitting, $N_H$ was  fixed to the quiescent state value along with the parameters of the first two temperature components.  Abundances and  temperature and normalization of the third component were free parameters in the spectral fitting. The resulting  temporal evolution of the spectral parameters of the flare is shown in Figure \ref{fig:para} and derived parameters  are given in Table \ref{tab:para}.

Abundances, temperature, and the corresponding emission measure  were found to vary during the flare. The peak value of $Z$ was derived to be $0.199 Z_\odot$, which is well above the 4$\sigma$ level to the minimum value observed. The flare temperature peaked at 72.8 MK during the rise phase of the flare, which is $\sim 3$ times more than the minimum value at the end of the decay phase. The EM  followed the flare light curve and peaked later than  the temperature at a value of  $9.72\times10^{52}$ cm$^{-2}$, which is $\sim 9$ times more than the minimum value observed at the end of the flare. The \lx ~  reached a value of $3.54\times10^{30}$ \lum, which is 1.8 times more than that of the quiescent state. The flux during  "U'' was 1.3  times more than that during quiescent state.

\subsubsection{Density measurement: The RGS spectra}\label{sec:density}

\begin{table*}
\centering
\caption{Best Fit Spectral Parameters from Different Segment of the Light Curve.}\label{tab:para}
\begin{tabular}{p{1.1in}p{1.1in}p{1.1in}p{1.1in}p{1.1in}p{0.6in}}
\hline
Time bins (ks) & $Z$          & T              & EM                 & $L_X$ & $\chi_\nu^2$(dof)\\
(from-to) & $(Z_\odot)$  &($10^7$ \deg K) &($10^{52} ~cm^{-3}$)& $~(10^{30}$ \lum) & \\
\hline
  6.0-11.0 &$0.148_{-0.003}^{+0.003}$   &$5.70_{-1.13}^{+1.57}$   &$1.94_{-0.26}^{+0.29}$  &$ 2.29_{-0.02}^{+0.02}$&1.26(361)\\
 11.1-12.2 &$0.138_{-0.007}^{+0.007}$   &$4.68_{-0.00}^{+0.00}$   &$1.98_{-0.46}^{+0.54}$  &$ 2.24_{-0.06}^{+0.06}$&1.10(223)\\
 12.2-13.0 &$0.166_{-0.009}^{+0.009}$   &$7.28_{-1.37}^{+1.38}$   &$6.18_{-0.73}^{+0.81}$  &$ 3.12_{-0.07}^{+0.07}$&1.00(220)\\
 13.0-13.4 &$0.170_{-0.008}^{+0.008}$   &$5.13_{-0.54}^{+0.62}$   &$8.74_{-0.74}^{+0.82}$  &$ 3.40_{-0.06}^{+0.06}$&1.02(272)\\
 13.4-14.0 &$0.199_{-0.012}^{+0.013}$   &$3.68_{-0.38}^{+0.66}$   &$9.72_{-0.59}^{+0.84}$  &$ 3.54_{-0.08}^{+0.08}$&1.11(203)\\
 14.0-15.0 &$0.179_{-0.014}^{+0.009}$   &$2.55_{-0.28}^{+0.44}$   &$7.60_{-0.46}^{+0.76}$  &$ 3.01_{-0.06}^{+0.06}$&1.11(241)\\
 15.0-17.0 &$0.170_{-0.007}^{+0.006}$   &$2.70_{-0.38}^{+0.43}$   &$4.31_{-0.51}^{+0.39}$  &$ 2.61_{-0.04}^{+0.04}$&1.19(282)\\
 17.0-30.0 &$0.170_{-0.002}^{+0.002}$   &$3.40_{-0.34}^{+0.33}$   &$2.81_{-0.21}^{+0.21}$  &$ 2.49_{-0.01}^{+0.01}$&1.08(487)\\
 30.0-37.0 &$0.166_{-0.003}^{+0.003}$   &$4.64_{-1.19}^{+2.13}$   &$1.12_{-0.24}^{+0.20}$  &$ 2.28_{-0.02}^{+0.02}$&1.12(386)\\
\hline
\end{tabular}
\end{table*}

\begin{figure*}
\begin{center}
\subfigure[{\sc Ovii} triplet]{\includegraphics[angle=-90,width=8.5cm]{fig4a.ps}}
\subfigure[Density]{\includegraphics[angle=-90,width=8.50cm]{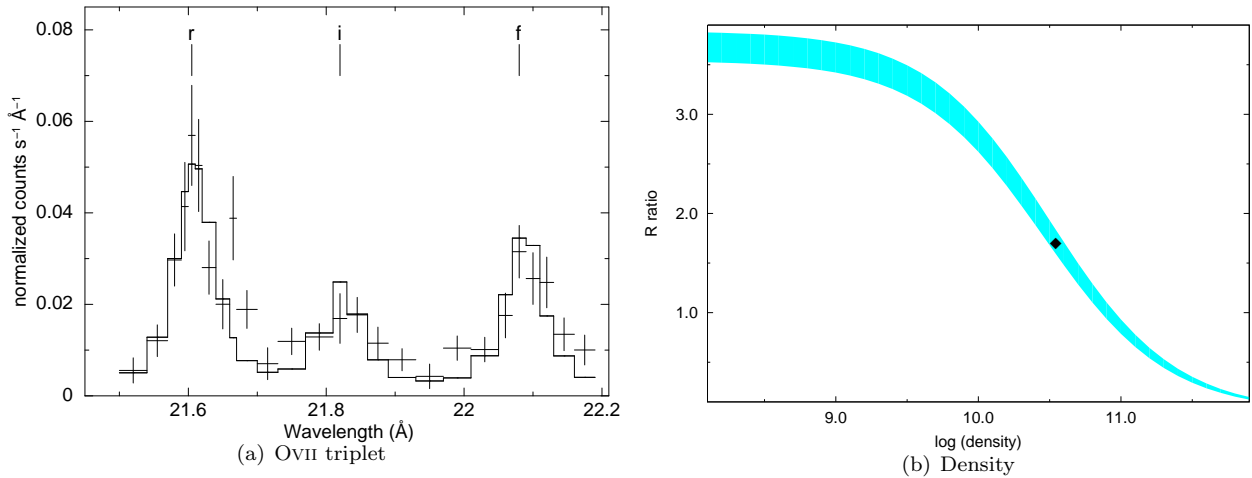}}
\caption{(a) He-like triplet of O{\sc vii}  extracted from the \xmm ~RGS spectrum of 47 Cas, (b) the  intensity ratio R = f/i of the O{\sc vii} forbidden  and intercombination  lines as a function of electron density, calculated using the CHIANTI database. }
\label{fig:rif}
\end{center}
\end{figure*}

The RGS spectra of 47 Cas were analyzed in detail by \cite{ness03,telleschi05} and \cite{nordon08}; therefore, we have restricted our analysis to the O{\sc vii} line only to  determine the density.  The density values during  the flare and the quiescent state of 47 Cas were derived by using He-like triplets from O{\sc vii}. The most intense He-like lines correspond to transitions between the $n = 2$ shell and the $n = 1$ ground state shell. The excited state transitions $^1$P$_1$, $^3$P$_1$ and $^3$S$_1$ to the ground state $^1$S$_0$ are called resonance ($r$), intercombination ($i$) and forbidden ($f$) lines, respectively. In the X-ray spectra, the ratio of fluxes in forbidden  and intercombination lines ($R = f/i$) is potentially sensitive to density ($n_e$), while the  ratio $G = (f+i)/r$ is sensitive to temperature \citep{porquet01,gabriel69}.  Of the He-like ions observed with the RGS,  O{\sc vii} has lines that are  strong and unblended to use in a measurement of $n_e$.  Figure \ref{fig:rif} (a) shows the He-like triplet from the O{\sc vii} line. Line fluxes  were measured using the {\sc xspec} package by fitting the RGS1  spectra with a sum of narrow Gaussian emission lines convolved with the response matrices of the RGS instruments. The continuum emission was described using Bremsstrahlung models at the temperatures of the plasma components inferred from the analysis of the EPIC spectra during the quiescent state.    We have used the CHIANTI atomic  database version 7.1 \citep{dere97, landi13} to derive the $G-$ and $R-$ratios.  The $G-$ratio of $\sim 1.0\pm0.2 $ for 47 Cas  implies a formation temperature of the {\sc Ovii} triplet of  $\sim 2\pm1$ MK, which is similar to that for other  active stars \citep{ness02}. Figure   \ref{fig:rif} (b) is a plot of the $R-$ratio and density as calculated from the CHIANTI atomic database.  A value of the $R-$ratio of  $1.7\pm0.4$ leads to an electron density of $4.0_{-1.5}^{+2.3}\times10^{10}$ cm$^{-3}$  during the flare. However, the electron density during the quiescent state was estimated as $2.5_{-1.7}^{+2.1}\times10^{10}$ cm$^{-3}$, which is well withinthe  $1\sigma$ level of the electron density during the flare. 

\section{Loop length and flare parameters}
\label{sec:loop}
Flares cannot be resolved spatially on a star, but show many analogies with solar flares.  Therefore, by analogy with solar flares and using flare loop models, it is possible to infer the physical size and morphology of loop structures involved in a stellar flare.  In order to derive loop parameters, many approaches have been developed for the analysis of stellar flares; in general, these are  based on some physical model of the flaring region and a fit of the observed decay behavior to the model.  \cite{reale97} introduced a time-dependent hydrodynamic model to infer the geometrical size of the flaring loop, which is  based on the decay time, and  the evolution of temperature and  EM during the flare decay.  This model includes both plasma cooling and the effect of heating during flare decay. The loop length ($L$) from hydrodynamic method is given as

\begin{equation}
L ~(~cm~)~ = 5.4\times10^3    ~\tau_d ~ T_{max}^{1/2} ~F(\zeta)^{-1} ~~~~~~ {\rm for  ~0.35 <\zeta < 1.6}
\label{eq:lhd}
\end{equation}

\begin{figure}
\includegraphics[angle=-90,width=80mm]{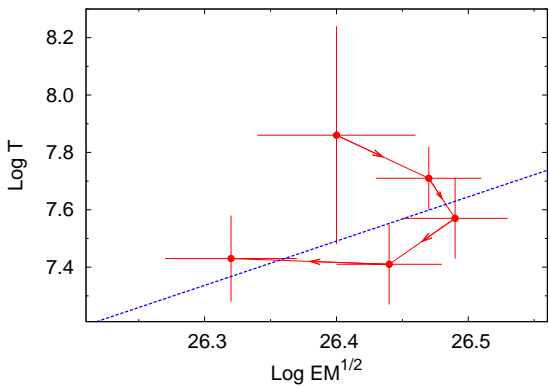}
\caption{Evolution of flare in log $\sqrt{\rm EM}$ \textendash ~log T plane. The dashed line show the best fit straight line during the decay phase with slope $\zeta$ = $1.54\pm1.06$. }
\label{fig:ktem}
\end{figure}

\noindent
The observed peak temperature must be corrected to a maximum value  using $T_{max} = 0.13 T^{1.16}$, where $T$ is the maximum best-fit temperature derived from spectral fitting to the data. The unitless correction factor is  $F(\zeta) = \frac{0.51}{\zeta-0.35}+1.35$; where $\zeta$ is the slope of the log $\sqrt{\rm EM}$ \textendash ~log $T$ diagram \citep{reale07}. Figure \ref{fig:ktem} shows the  log $\sqrt{\rm EM}$ \textendash ~log T diagram during the flare. We derive the value of $\zeta$ to be $1.54\pm1.06$, indicating the presence of sustained heating during the decay of the flare was negligible.  The value of $\zeta$ in the upper extreme is outside the domain of the validity of the method, therefore, the loop length was derived by using the lower extreme value of $\zeta$ (i.e. 0.48) and estimated as $ 3.3\times10^{10}$ cm.

After finding $L$, $EM$ and $n_e$, we can derive the pressure ($p$), the volume ($V$), and the minimum magnetic field ($B$) to confine the flaring plasma as 

\begin{eqnarray}
p = 2 n_e ~k ~T  ~~{\rm dyne ~cm^{-2}};~~~
V = EM ~ n_e^{-2} ~~{\rm cm^3};~~~ \nonumber \\
B = \sqrt{8\pi ~p}  ~~{\rm G}
\end{eqnarray}

\noindent
using $n_e = 4.0\times10^{10}$ cm$^{-3}$, $EM = 9.72\times10^{52}$ cm$^{-3}$ and $T = 7.28\times10^7$ K. The estimated values of  $p$, $V$ and $B$ during the flare observed in 47 Cas  were 804 dyne cm$^{-2}$, $6.0\times10^{31}$ cm$^3$ and 142 Gauss, respectively. 


\section{Discussion and conclusions}
\label{sec:disc}
We have carried out an analysis of the flare observed in 47 Cas  using  X-ray data from \xmm. The rise and decay times of the flare  were found to be similar to those of the flares observed  from G-K dwarfs \citep{pandey08} and are smaller than those from evolved RS CVn-type and pre-main-sequence stars \citep{pandey12,getman08a}. These data along with the flare luminosity suggest that the flare from 47 Cas is an impulsive flare \citep{pallavicini90}, which is similar to a compact solar flare. The compact flares are less energetic ($\sim 10^{30}$), short in duration ($<1$ hour) and confined to a single loop. The \td ~in the soft energy band was higher  than that in the hard energy band. A similar tendency was also noticed in many flares observed from G-K dwarfs.  This could probably be due to the softening of the spectrum during the decay due to the plasma cooling, i.e. emission gradually shifts from the high energy band to deeper in the soft energy band.  We have also noticed the heightened emission after the flare in the soft energy band. Similar behavior in light curves  was seen in the flares from $\xi$ Boo \citep{pandey09} and CC Eri \citep{pandey08,chacon07}.  It also appears that before the flaring event the X-ray light curve was not constant.  Such small-scale variability in the X-ray light curve before and after the flaring event could be due to the emergence of smaller flares during the observations.  Continuous low-level variability due to small flares  has also been reported for active dwarfs and giants \citep{vilhu93,kuerster97,mathioudakis99,ayres01}.  We have derived a maximum temperature of 72.8 MK at the beginning of the flare, which is more than that observed in many flares from similar dwarfs  \citep{pandey08}. The EM  peaked later than the temperature during the decay of the flare. Similar delay often has been observed both in solar and stellar flares \citep{sylwester93, favata00, maggio00, stelzer02, pandey08,pandey12}. This could be due to a coherent plasma evolution during the flare and therefore, a flare occurring inside a single loop, or at least the presence of a dominant loop early in the flare. We found a significant increase in the abundances from 0.13 \zsun ~to 0.2 \zsun during the flare. A possible explanation for an enhancement in abundances during the  flare is due to the evaporation of  fresh chromospheric material in the flaring loops.  Using the RGS spectra of 47 Cas, \cite{nordon08} and \cite{telleschi05} found  a small enhancement of abundances of low first-ionization-potential metals. However, this enhancement in the abundances was well within a $1\sigma$ level.

The length of the flaring structure was derived to be $3.3\times10^{10}$ cm using a state-of-art hydrodynamic method based on the decay phase.  We have also derived the loop length from other methods as given by \cite{haisch83}, \cite{hawley95}, \cite{shibata02} and \cite{aschwanden08}, and we found that  the loop lengths from these methods are consistent with that from \cite{reale97}'s method. Many authors in the past have compared the loop lengths from the above methods and some of them found consistent loop lengths \citep{covino01,bhatt14}; however, others found inconsistencies in the loop length determination \citep{favata01,shibata02,srivastava13}.  Considering 47 Cas to be an early G-type star \citep{guedel98},   the loop length was found to be less than the half of the stellar  radius and much less than the pressure scale height of $\sim 4\times10^{11}$ cm. Here, pressure scale height is defined as $h= k T/\mu g$; where $k$ is the Boltzmann constant, $T$ is the plasma temperature, $\mu$ is the mean molecular weight in terms of the proton's rest mass, and $g$ is the surface gravity of the star.  The pressure and density derived for the present flare are intermediate between those of the flares from G-K dwarfs \citep{pandey08}.  

Due to the lack of multi-wavelength coverage, a detailed assessment of the energy budget of the present flare is not possible. However, using the scaling laws from \cite{rosner78}, the heating rate per unit volume can be determined as $E_H = 10^{-6} ~ T^{3.5} ~ L^{-2} ~~{\rm erg ~s^{-1} ~cm^{-3}}$, where $T$ and $L$ are plasma temperature and loop length. Using $L = 3.3\times10^{10}$ cm, the value of $E_H$ was obtained as $3.2 {\rm ~erg ~s^{-1} ~cm^{-3}}$.  The total heating rate at the flare maximum is estimated  as, $H = E_H V \approx 2\times10^{32} {\rm ~erg ~s^{-1}}$.  In order to satisfy the energy balance relation for the flaring as a whole, the maximum X-ray luminosity must be lower than the total energy rate ($H$) at the flare peak.  For the present flare, the total energy rate was found to be $\sim 40$ times more than the  peak X-ray luminosity. This value is in agreement with those reported for the solar flares where the soft X-ray radiation only accounts for up to 20 \% of the total energy \citep{wu86}, but is smaller than many flares from solar-like stars. The total energy released from the flare was found to be $2\times 10^{34}$ erg over 4.8 ks, which is  equivalent to $\sim 3$ s of the star's bolometric energy output. The total X-ray energy released during the flare from 47 Cas indicates that this flare was as energetic as flares from other G-K dwarfs ($2.3\times10^{32} - 6.1\times10^{34}$ ergs; Pandey \& Singh 2008).  If we assume heating is constant for the initial rise phase, which lasts for \tr  $\approx 831$ s, and then decays exponentially, with an e-folding time of \td $\approx 2494$  s, the total energy [$E_{tot} = H$(\td+\tr)] is estimated as $\sim 6\times 10^{35}$ erg, which is approximately 30 times more than the energy radiated in X-rays.

It is believed that the magnetic field provides the main source of energy for the solar/stellar  activities including flares.  In order to know the strength of the magnetic field ($B_m$) required to accumulate the emitted energy, we assume that the energy released during the flare is indeed of magnetic origin and it occurs entirely within a single coronal loop structure. The total energy can be estimated as

\begin{equation}
E_{tot} = \frac{(B_m^2-B^2)}{8 \pi} V
\end{equation}

\noindent 
Using the values of $E_{tot}$, $B$ and $V$, the total magnetic field required to produce the flare is estimated to be $\sim 517$ Gauss.  

The parameters derived from the present analyses indicate that the X-ray flare on 47 Cas is similar to a solar impulsive flare whose height ($L/\pi$) is only 15\% of the stellar radius.

\acknowledgments

This work uses data obtained by \xmm, an ESA science
mission with instruments and contributions directly funded by ESA
Member States and the USA (NASA). The authors also thank the referee for his/her comments and suggestions. Authors acknowledge the project reference INT/RUS/RFBR/P-167. 

{\it Facilities:} \facility{\xmm ~}, \facility{SAS}

\end{document}